\input harvmac
%\draft

%\def\b{- \sqcup\!\!\!\!\!\!\!\;\sqcap}
%\def\b{- \partial^2}
\def\d{\;i\! \not\!\partial\;}
\def\sA{\not\!\! A}
\def\p{\not\! p}
\def\ld{(-i\! \not\!\partial^{\!\!\!\!^\leftarrow}\!)\;}
\def\rd{\;i\! \not\!\partial^{\!\!\!\!^\rightarrow}}
\def\lrd{\not\!\partial^{\!\!\!\!^\leftrightarrow}}
\def\lD{{\cal D}^{\!\!\!\!^\leftarrow}}
\def\rD{{\cal D}^{\!\!\!\!^\rightarrow}}
\def\cA{{\cal A}}
\def\cD{{\cal D}}
\def\lA{{\cal A}^{\!\!\!\!^\leftarrow}}
\def\rA{{\cal A}^{\!\!\!\!^\rightarrow}}
\def\ps{\bar\psi}
\def\bc{\bar c}
\def\chib{\bar\chi}
\def\pib{\bar\pi}
\def\phib{\bar\phi}

\def\h{{1 \over 2}}
\def\ih{{i \over 2}}
\def\og{{1\over g}}
\def\i{\int\!\!{\rm d}^3x\;}
\def\M{(m_1+m_2)}
\def\Mh{{m_1+m_2\over 2}}
\def\e{\epsilon}

\lref\fs{L.D.  Faddeev and A.A. Slavnov, {\it Gauge fields.
Introduction to quantum theory,} 2nd edition, Benjamin 1989.}
\lref\hwkng{S.W. Hawking, {\it Who's Afraid of (Higher Derivative)
Ghosts?}, In Quantum Field Theory and Quantum Statistics,
vol.2, p.129, ed. I.A. Batalin et al Cambridge, 1985.}
\lref\www{see, e.g., A. Bartoli and J. Julve, {\it Nucl. Phys.} {\bf B425}
(1994) 277 and references threin.}
\lref\brst{T. Kugo and S. Uehara, {\it Nucl. Phys.} {\bf B197}
(1982) 378 \semi
L. Baulieu, {\it Phys. Rep.} {\bf 129} (1985) 1.}
\lref\colect{A. Hosoya and K. Kikkawa, {\it Nucl. Phys.} {\bf B101}
(1975) 271 \semi
J. Alfaro and P.H.Damgaard, {\it Ann. Phys.} (N.Y.)
{\bf 202} (1990) 398.}
\lref\mghost{A. Salam and J. Stathdee, {\it Phys. Rev.} {\bf D18}
(1978) 4480.}

\Title{\vbox{\baselineskip12pt\hbox{INRNE-TH-97/6}}}
{\vbox{\centerline{On the Connection between Pauli---Villars}
\vskip2pt\centerline{and Higher Derivative Regularizations}}}
%   \footnote{}{*optional footnote on title}

\centerline{M.N. Stoilov\footnote{*}{
e-mail: mstoilov@inrne.acad.bg}
\footnote{$^\star$}
{Supported in part by Bulgarian National Scientific Foundation
 under contract $\Phi$-401-94}
}

\bigskip\centerline{Institute for Nuclear Research and Nuclear Energy}
\centerline{Boul. Tzarigradsko Chaussee 72, 1784 Sofia, Bulgaria}

\vskip .3in

\noindent{\bf Abstract. }
We show that in some cases the gauge invariant Pauli---Villars
and higher (covariant) derivatives regularizations are equivalent.

\Date{10 June 97}

Higher covariant derivatives and gauge invariant Pauli---Villars
regularizations have a quite special place
in the long list of regularizations used in Quantum Field Theory.
First, a combination of these regularizations is used to prove the
renormalizability of Yang---Mills theories \fs .
Second, they are the only ones which could be
incorporated into the Lagrangian of the model as additional local terms.
In this paper our aim is to show that these regularizations have
something more in common --- in fact, in some cases they are just two
different forms of a same regularization.

It seems that higher derivatives (HD) regularization originates from the
usual Pauli---Villars (PV) one.
The latter prescribes to replace in loop calculations the propagator
$(\p - m)^{-1}$
with
$
(\p+m)\left((p^2-m^2)^{-1}+\sum_{j=1}^k c_j(p^2 - m_j^2)^{-1}\right).
$
(Here we deal with spinors only.
However, the same approach could be applied to field
with any spin and statistics.)
The constants $c_j$ and $m_j$ are such that
\eqn\cond{
1 + \sum_{j=1}^k c_j = 0, \;\;\;
m^2 + \sum_{j=1}^k c_j m_j^2 = 0, \;\;\;
{\rm etc.} \;\;  .
}
Equations \cond\ allow $c_j$ to be integer.
In this case the regularized propagator could be put in the form
%(provided that and $m_i \neq m_j$ for $i\neq j$)
$
{1\over g}(\p - m)^{-1} \prod_{j=1}^k(p^2 - m_j^2)^{-1},
$
where
$
(-1)^k g^{-1}=\prod_{i=1}^k m_i^2+\sum_j^k c_j m^2 \prod_{i\neq j}m_i^2.
$
On the other hand this propagator could be viewed as obtained from
a Lagrangian with the following free term
$
L^{free} = g \i\ps (\d - m) \prod_{j=1}^k (-\partial^2 - m_j^2)\psi.
$
Replacing this particular form of $L^{free}$ with the most general
(polynomial) expression and usual derivatives with covariant ones one
obtains a (variant of) higher covariant derivatives regularization for
spinor field. The spinor part of the Lagrangian in this case is
(here $g$ is a constant with dimension $mass^{-k}$
and $A$ is the gauge potential)
\eqn\HcD{
L = g\i\ps(\d+\sA-m)\prod_{j=1}^{k}(\d+\sA-m_j)\psi.
}
%

%This is the HD Lagrangian we shall compare with the gauge invariant PV
%regularized spinor Lagrangian
One of the possible viewpoints to the gauge invariant PV regularization
is that in divergent diagrams
one regularises a whole spinor loop at a time adding and subtracting
the same diagram (with some integer coefficients $c_i$) but with
different masses in the propagators forming the loop.
It is possible to write down a Lagrangian which reproduce
automatically this scheme and it has the form
\eqn\giPV{
L =\int \ps(\d+\sA-m)\psi +
\sum_i\sum_{j=1}^{c_i}\ps_{ij}(\d+\sA-m_i)\psi_{ij} +
\sum_k\sum_{j=1}^{\vert c_k\vert}\phib_{kj}(\d+\sA-m_k)\phi_{kj}.
}
The sum in the second term in  \giPV\ is over positive coefficients $c$
and so, the extra fields $\psi_{ij}$ are with Fermi statistics;
the sum in the third term is over negative coefficients and thus
$\phi_{kj}$ are Bose fields.
Usually, extra fields have one and the same mass $M$.
This has some advantages, however, using different masses one can fix
$\vert c_i\vert = 1$.
In this case \giPV\ takes the form
\eqn\giPVp{
L = \i\ps(\d+\sA-m)\psi +
\sum_i\ps_i(\d+\sA-m_i)\psi_i +
\sum_k\phib_k(\d+\sA-m_k)\phi_k.
}

Our goal in this letter is to show that model with Lagrangian \HcD\ is
equivalent (in some cases) to that with Lagrangian \giPVp .
The first step is to transform \HcD\ into first order Lagrangian.
After that we represent the HD fermionic ghosts \hwkng\ arising in step one
as boson ghosts.
A short note is needed at this stage before go further.
The extra fields in the gauge invariant PV regularized Lagrangians
\giPV\ and \giPVp\ are unphysical and {\it a priori}
there are not sources for them in the generating functional of the model
\fs .
Something similar has to take place in the models with HD ---
it is easy to show that
provided $m_i \neq m_j \;\;\;\forall\;\; i\neq j$
the general solution of a HD equation
$\prod_j(\d - m_j)\psi = 0$ is
$\psi = \sum_j\psi_j$ where $(\d- m_j)\psi_j =0$.
Therefore, after quantization $\psi$ describes a set of ordinary fields.
However, only one of these fields is physical and so, there has to be
source only for it.
(This is obvious if one looks at HD regularization as a variant of the
usual PV one.)
A possible way to achieve this is to enforce 'by hands'
$(\d - m)\vert {\rm ph}> = 0$; another possibility is discussed in
\mghost .
In fact it does not matter for us how the problem is cured.
The only important thing is that there are not sources for extra fields
in HD case too.
As a consequence, we can work simply with the Lagrangians \HcD\ and
\giPVp\ and not with corresponding generating functionals.

We begin our considerations on the conversion of HD to first order
Lagrangian with a simple example of second order (in
derivatives) free Lagrangian for spinor field
\eqn\second{\eqalign{
L'' & = g \i \ps (\d - m_1)(\d-m_2) \psi \cr
 & = g \i \ps\ld(\rd)\psi - {g\over 2}\M\ps\lrd\psi +
gm_1m_2\ps\psi.     \cr
}}
As it was mentioned above if $m_1 \neq m_2$ (for definitness we use $m_1>m_2$)
the solution of the equation of motion is
$\psi = \psi_1+\psi_2,$ where $(\d-m_i)\psi_i = 0$.
Moreover, any dynamical invariant (energy-momentum, charge, etc.) is a sum
of the corresponding invariants for usual spinor fields with mass $m_1$
and $m_2$ (one of them with minus sign).
These facts suggest that $L$ itself also could be presented as a sum of
usual fermionic Lagrangians.
We shall demonstrate this using suitable Legendre transformation.
The procedure is an analogue of the one used for Lagrangian derivation
of Hamiltonian equations and is often used in the analysis of HD theories
\www .
Let us introduce the quantities
$$%
%\eqn\varbls{
\eqalign{
a &\equiv \rd \psi \cr
\bar a &\equiv \ps \ld . \cr
}%}
$$%
The functional variations of $L$ with respect to $a$ and $\bar a$ are
$$%
%\eqn\momenta{
\eqalign{
\pi \equiv {\delta L'' \over \delta \bar a}& =
g \ps\ld -{g\over 2} \M\ps \cr
\bar \pi \equiv {\delta L'' \over \delta a}& =
g \rd\psi -{g\over 2} \M\psi . \cr
}%}
$$%
These identities are used to express $a$ as a function of $\pi$
and $\psi$.
The Legendre transform of $L''$ with respect to $a$ and $\bar a$ is
$$%
%\eqn\ham{
\eqalign{
\Lambda\left[\pib,\ps,\pi,\psi\right] &=
\i (\pib a + \bar a\pi) - L''\cr & =
\i  g(\og\pib +\Mh\ps)
(\og\pi+\Mh\psi)-gm_1m_2\ps\psi \cr
}%}
$$%
and the first order Lagrangian governing the dynamics of our model is
\eqn\first{\eqalign{
L'& = \i \pib\rd\psi + \ps\ld\pi - \Lambda \cr
  & = \i \pib\rd\psi + \ps\ld\pi -
\og\pib\pi -\Mh(\pib\psi +\ps\pi) -
{g\over 4}(m_1-m_2)^2\ps\psi.  \cr
}}
It is easy to check that the equations of motion for $\psi$ and $\ps$,
following from \first\ coincide with those from \second .
Now we would like to diagonalize  $L'$.
For this purpose we introduce the linear combination
$$ {\phi\choose\chi} =  U{\pi\choose\psi} $$
where $U$ is some $2\times 2$ complex matrix such that
$\vert {\rm det}U\vert^2 = 1$.
Fixing the elements $u_{ij}$ of the $U$ so that
$$%
%\eqn\fix{
\eqalign{
u_{11} & = \pm \h g (m_1-m_2) u_{21} \cr
u_{12} & = \mp \h g (m_1-m_2) u_{22} \cr
\vert u_{21}\vert^2 & = 1/\vert g (m_1-m_2)\vert =
\vert u_{22}\vert^2 \cr
}%}
$$%
$L'$ takes the form
\eqn\diag{L' = {g\over\vert g\vert} \i
\ih\phib\lrd\phi - m_1\phib\phi -
\ih\chib\lrd\chi + m_2\chib\chi.
}

We see that the our initial second order theory can be described by
a difference of two usual spinor Lagrangians for
two (not interacting) Fermi fields.
The field $\chi$ which Lagrangian enters \diag\ with minus sign
is a ghost field \hwkng .
It is possible to change this bad sign in the kinetic term
but as we shall see this is of little use.
The change can be achieved by a suitable antiunitary transformation
(of the time-reverse type) applied on $\chi$ field.
Namely, let us denote with $'$ the quantities after transformation
(as usual, the antiunitary transformed of any operator $A$ is
$A' = (U^{-1}AU)^\dagger$).
For $\chi$ field we have:
$$%
%\eqn\trans{
\eqalign{
\chi' & = \eta\chib T \cr
\chib' & = \eta^* T^{-1}\chi,\;\;\;\vert\eta\vert^2 = 1.\cr
}%}
$$%
Here $T$ is some matrix, we want to satisfy $T^\dagger = T^{-1}$ and
$\gamma^0T\gamma^0 = T$. The kinetic term for $\chi$ field changes the
sign provided that in additional
\eqn\signch{
T\gamma^{\mu\!\!\!\!\!\!\!^{\rm T}}\;T^{-1}=\gamma^\mu.
}
The matrices $\gamma^{\mu\!\!\!\!\!\!\!^{\rm T}}\;$
satisfy the identities for $\gamma$-matices, so
there is a (unitary) matrix $T$ with desired property \signch .
Thus, formally, we can write $L'$ as a sum of two ordinary spinor
Lagrangians but with different signs of the mass terms.
However, the transformation used is not unitary and therefore, on
quantum level the two theories will differ.
If we want to keep the contact with higher derivative theory, we
should quantize one of the fields in a non standard way thus coming back
to the form \diag\ of $L'$.

Now we proceed the $2n$-th order case in the presence of gauge interaction.
It is always possible to write its spinor Lagrangian in the form
\eqn\sno{
L^{(2n)} =
g \i\ps\lD(\rD + \rA)\psi
}
where
$$%
%\eqn\dfn{
\eqalign{
\rD & = \prod_{j=1}^n(\rd + \sA - m_j), \cr
%\;\;\; m_i\neq m_j \;\; \forall i\neq j\cr
\lD & = \prod_{j=1}^n(\ld + \sA - m_j),\cr
}%}
$$%
and $\cA$ is some operator of order $k<n$ which commutes with $\cD$.
As a consequence every function of $\cA$ and $\cA^{-1}$
(which should be understand as a power series of $\cA$)
commutes also with $\cD$.
The variables we use in Legendre transformation are
$$%
%\eqn\var{
a \equiv \rD\psi,\;\;\; \bar a \equiv \ps\lD;
%}
$$%
the corresponding momenta are
$$%
%\eqn\mom{
\pib = g(\bar a + \h\ps\lA),\;\;\; \pi = g(a + \h\rA\psi),
%}
$$%
and the $n$-th order Lagrangian, equivalent to $L^{(2n)}$ is
\eqn\no{
L^{(n)} =
\i\pib\rD\psi + \ps\lD\pi -
g(\og\pib - \h\ps\lA)(\og\pi - \h\rA\psi)
}
Again, as in the second order case,
the equations of motion for $\psi$ and $\ps$ following
from \no\ reproduce the ones from \sno .

Up to this point we simply repeat the procedure used in the second
order case.
The not so straightforward step is the diagonalization of the $L^{(n)}$.
Consider the following variables' change
\eqn\ole{
\eqalign{
\chi & = \sqrt{{g\over\rA}}(\og\pi + \h\rA\psi), \cr
\phi & = \sqrt{{g\over\rA}}(\og\pi - \h\rA\psi), \cr
}}
and analogously for $\chib$ and $\phib$.
The Berezian of this change of variables is $1$ and so,
\ole\ leaves the integration measure in the generating functional invariant.
In the new variables $L^{(n)}$ takes the form
\eqn\nof{
L^{(n)} =
\chib\rD\chi - \phib(\rD + \rA)\phi.
}
Note, that deriving \nof\ we have not used the fact that $k$ (the
order of $\cA$) is less than $n$ (the order of $\cD$).
The only thing that is really important is that $k\neq n$.
Therefore, we could consider the case $k = n+1$ and thus
to cover all HD Lagrangians.

Repeating the above steps sufficiently times we could present the HD
Lagrangian \HcD\ of arbitrary order $n$ as a sum of $n$ first
order Lagrangians (with altering signs) for $n$ independent Fermi
fields
\eqn\linear{
L = \i \sum_i \ps_i(\d + \sA -m_i)\psi_i -
       \sum_i \chib_i(\d + \sA -m_i')\chi_i.
}
Same is the structure of the gauge invariant PV regularized
Lagrangian.
The only difference is that in \giPVp\ there are not spinor terms with opposite
signs, but Bose terms (with correct sign).
Now we want to show that this difference could be removed.
For this we use the so called 'collective field method' \colect\
 --- we introduce extra gauge freedom in the model and then fix it.
In fact, we fix the gauge in two different ways.
The first gives the linear decomposition \linear\ of the HD
Lagrangian, the second - the gauge invariant PV regularization.
To clarify the idea let us consider a simple example with only one
(spinor) field and Lagrangian
\eqn\sempl{L = \i\ps\cD\psi.}
Here $\cD$ is some operator we shall not specify %yet
and we suppose there are no sources for $\psi$ and $\ps$ in the
generating functional of the theory.
Let us introduce extra (collective) field $\phi$, so that the Lagrangian
becomes
\eqn\extra{L = (\ps +\phib)\cD(\psi + \phi).}
This Lagrangian possesses an extra local symmetry
\eqn\es{\eqalign{
\delta\psi &= -\rho \cr \delta\phi &=\rho,\cr}}
where $\rho$ is an arbitrary spinor function.
Following \brst\ we introduce an auxiliary field $\lambda$ and a
ghost pair $\{c,\bc\}$ for this gauge symmetry
(the ghosts are bosons due to the spinor character of the $\rho$).
After gauge fixing the Lagrangian is invariant under rigid BRST
symmetry.
The infinitesimal BRST transformation of the fields we are interested in
are (here $\e$ is the parameter of the transformation)
$$
\eqalign{
\delta_Q\psi &= -c\e, \cr
\delta_Q\phi &= c \e,\cr
\delta_Q\bc & = \lambda, \cr
\delta_Q\lambda &= 0. \cr
}
$$
The BRST invariant Lagrangian has the form  \brst\
$$
L_{BRST} = L + \delta_Q(\bc\varphi),
$$
where $\varphi$ is the gauge fixing condition and $L$ is that of formula
\extra .
Choosing $\varphi=\phi$ we get
$$
\delta_Q(\bc\varphi) = \lambda\phi + \bc c
$$
Thus the ghosts trivially decouple from the dynamics of the system, the
field $\phi$ is set to zero and we restore the initial model with
Lagrangian \sempl .
If we choose $\varphi = \cD\phi$ the result reads
\eqn\one{
L_{BRST} = \ps\cD\psi + \bc\cD c .
}
The same result is obtained if we consider
instead of gauge transformation \es\ the following one
$$%
%\eqn\esd{
\eqalign{
\delta\psi &= -\cD\rho \cr \delta\phi &=\cD\rho\cr}
%}
$$%
with gauge condition $\varphi=\phi$.
%In what follows we  use intensively transformations of this type.
%
In our next step we introduce the collective field in a slightly
different way.
We replace \sempl\ with
\eqn\xxtra{
L = \i(\ps +\phib \cA)\cD(\psi +\cA\phi),}
where $\cA = \sqrt{\cD^{-1}}$. ($\cA$ should be understand again
as a series of $\cD$).
The extra symmetry in this case is
$$%
%\eqn\xs{
\eqalign{
\delta\psi &= -\sqrt{\cD}\rho \cr \delta\phi &=\cD\rho,\cr}
%}
$$%
with obvious BRST symmetry of the fields.
Choosing gauge fixing function to be $\varphi = \phi$ we get
$$
 L_{BRST} = \ps\cD\psi + \bc\cD c
$$
which coincides exactly with \one .
The alternative choice $\varphi = \sqrt{\cD}\psi$ leads to
$
 L_{BRST} =\i \phib\phi - \bc\cD c.
$
The field $\phi$ is therefore nondynamical and we are left with
\eqn\two{ L_{BRST} =  - \bc\cD c }
The sequence of equivalences between \sempl , \one\ and \two\
shows that the dynamics of Fermi field with Lagrangian
\sempl\ is equivalent to  the dynamics of Bose field with Lagrangian
\two\ if there are not sources for this field.
Applying this result to the terms in \linear\
with minus sign we prove the equivalence between higher
derivative and gauge invariant PV regularizations for a spinor field
in the case when $m_i \neq m_j \;\;\;\forall\;\; i\neq j$.

\bigskip
\bigskip
\centerline{\bf Acknowledgements}

We thank BNSF contract $\Phi-401-94$ for the
yearly support of US dollars 50.

\listrefs
\bye

\end